\documentstyle [11pt]{article}
\input{psfig}
\begin{document}
\title{Numerical Investigation of Cosmological Singularities \\
{\small}}
\author{Beverly K. Berger\\
Physics Department \\
Oakland University \\
Rochester, MI 48309 USA}
\maketitle
\begin{abstract}
A primary unresolved issue for cosmological singularities is whether or
not their behavior is locally of the Mixmaster type (as conjectured by
Belinskii, Khalatnikov, and Lifshitz (BKL)). The Mixmaster dynamics
first appears in spatially homogeneous cosmologies of Bianchi Types VIII
and IX. A multiple of the spatial scalar curvature acts as a closed potential
leading, in the evolution toward the singularity (say $\tau \to \infty $), to
an (almost certainly) infinite sequence of bounces whose parameters
exhibit the sensitivity to initial conditions usually associated with chaos.
Other homogeneous cosmologies are characterized by open (or no)
potentials leading to a last bounce as $\tau \to \infty$ . Such models are
called asymptotically velocity term dominated (AVTD). Here we shall
describe a numerical approach to address the BKL conjecture. Starting
with a symplectic numerical method ideally suited to this problem, we
shall consider application of the method to three models of increasing
complexity. The first application is to spatially homogeneous (vacuum)
Mixmaster cosmologies where we compare the symplectic ODE solver to
a Runge-Kutta one.
The second application is to the (plane symmetric, vacuum)
Gowdy universe on $T^3 \times R$. The dynamical degrees of freedom
satisfy nonlinearly coupled PDE's in one spatial dimension and time.
We demonstrate support for
conjectured AVTD behavior for this model and explain its observed
nonlinear small scale spatial structure. Finally, we study $U(1)$
symmetric, vacuum
cosmologies on $T^3 \times R$. These are the simplest spatially
inhomogeneous universes in which local Mixmaster dynamics is allowed.
The Gowdy code is easily generalized to this model,
although, the spatial differencing needed in the symplectic method is not
trivial.For AVTD models, we
expect the potential-like term in the Hamiltonian constraint to vanish
as $\tau \to \infty$ while in local Mixmaster it
becomes (locally) large from time to time. We show how the
potential behaves for a variety of generic $U(1)$ models.
\end{abstract}

\section{Introduction}
In these lectures, I propose to discuss the application of symplectic
numerical methods \cite{fleck,vm83}
to the investigation of cosmological singularities.  For
a system whose evolution can be described by a Hamiltonian, the
symplectic approach splits the Hamiltonian into kinetic and potential
subhamiltonians.  If the subhamiltonians are exactly solvable, these
solutions can be used to evolve the system from one time to the next.
Fortunately, an appropriate choice of variables in the standard $3 + 1$
Hamiltonian formulation of general relativity enables Einstein's equations
to be derived from a Hamiltonian appropriate for the symplectic algorithm
(SA).  So far, the primary application of SA to general relativity has been
to determine the asymptotic singularity behavior of cosmological models
(but see also \cite{sarwphi}). The SA is well-suited to this problem
because it becomes exact if the asymptotic behavior is asymptotically
velocity term dominated (AVTD)---i.e.~the kinetic subhamiltonian
asymptotically determines the dynamics.

\begin{figure}[bth]
\begin{center}
\setlength{\unitlength}{1cm}
\makebox[11.7cm]{\psfig{file=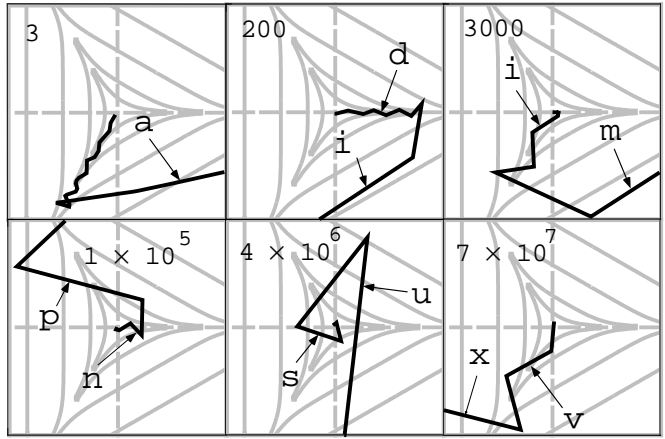,width=9cm}}
\caption [Typical Mixmaster Trajectory]
{\protect \label{mixtraj}
A typical Mixmaster trajectory in the anisotropy
plane with horizontal axis $\beta_+$ and vertical axis
$\beta_-$ which are centered on zero.  Time increases (and the
singularity is approached) to the right and downward.  The
number on each frame indicates the axes' scales.  The letters
label the Kasner epochs.  The Mixmaster minisuperspace
equipotentials are shown shaded in gray.}
\end{center}
\end{figure}

To study the application of SA to general relativity, we shall consider a
sequence of models whose variables depend on 0, 1, and 2 spatial
dimensions.  The first case we shall consider is the spatially homogeneous
Mixmaster universe.
(For convenience, we shall consider only the
diagonal Bianchi IX vacuum model.)  Einstein's
equations can be obtained by variation of the Hamitonian
\begin{equation}
\label{mixh0}
2 {\cal H} =-p_\Omega ^2+p_+^2+p_-^2+U(\Omega ,\beta _+,\beta _-)
\end{equation}
where
\begin{eqnarray}
\label{mixu}
U&=&e^{4\Omega }\left( {e^{-8\beta _+}+e^{4(\beta
_++\sqrt 3\beta _-)}+e^{4(\beta _+-\sqrt 3\beta _-)}}
\right.\nonumber \\
 & &\quad\quad\quad\quad\left. {-2e^{4\beta _+}-2e^{-2(\beta
_++\sqrt 3\beta _-)}-2e^{-2(\beta _+-\sqrt 3\beta _-)}}
\right).
\end{eqnarray}
Here $\Omega(t)$ is the logarithm of the cosmological scale factor and
measures isotropic expansion, while $\beta_{\pm}$ measure orthogonal
anisotropic shears with $p_{\Omega}$ and $p_{\pm}$
respectively canonically conjugate to $\Omega$ and $\beta_{\pm}$ \cite{mtw}.
The
potential $U$ is proportional to the spatial scalar curvature ($U = e^{6
\Omega} \ ^3\/\/R$) and is shown in Fig.~\ref{mixtraj}.
Eq.~(\ref{mixh0}) is itself the
Hamiltonian constraint ${\cal H} = 0$.  The properties of this model have
been known (more or less) since the late sixties \cite{BKL,misner}.  The
first three terms in $U$ describe triangular walls.  As the model
evolves toward the singularity at $\Omega = \infty$, the walls become
exponentially steep.  Within the walls, the system behaves almost
as a free particle
($U \approx 0$).  In the approach to the singularity, $\Omega$ itself may
be used as the time variable.  As $\Omega \to - \infty$, a fixed value of
$U$, say $U_0$, moves outward in the $\beta_{\pm}$-plane at a speed
$1/2$ that of the system point \cite{ryan}.  Thus the system evolves with
an infinite sequence of bounces off the potential.  A typical trajectory is
shown in Fig.~1.  While there is no exact solution, each straight line
segment can be parametrized and a map (the BKL map) derived to link
one segment to the next \cite{BKL,CB,bkb94}.  There is a long history
of numerical simulations \cite{moser}, \cite{hobillbook} trying to assess the
validity of the BKL map as a descriptor of the dynamics.  Part of this
interest can be traced to the fact that a bounce which leaves one of the
$120^{\circ}$ corners of the potential to move to another corner exhibits
the sensitivity to initial conditions usually associated with chaos.
Whether or not Mixmaster dynamics is chaotic remains a topic for both
analytic and numerical study \cite{hobillbook}.
We note that with $U=0$, the solution is that of Kasner \cite{Kasner}. A
solution which is asymptotically Kasner is AVTD.  The Mixmaster
solution is the antithesis of AVTD since there is (presumably) no last
bounce.  The effect of the spatial derivatives that generate $U$, though
almost absent during each Kasner epoch, always recurs.

Our remaining examples are spatially inhomogeneous cosmologies (still
vacuum for convenience).  Long ago, BKL conjectured
that the singularity
in spatially inhomogeneous cosmologies is locally of the Mixmaster type
\cite{BKL}.
Analytic verification of the BKL conjecture has bogged down on the issue
of setting up the local Mixmaster behavior in a global way.  For this
reason, a numerical approach may be useful.

While our second model's plane symmetry precludes local asymptotic
Mixmaster dynamics, it does serve as an excellent laboratory for the SA.
The Gowdy model
on $T^3 \times R$ is described by the metric \cite{gowdy}
\begin{eqnarray}
\label{gowdymetric}
ds^2&=&e^{{{-\lambda } \mathord{\left/ {\vphantom {{-\lambda
} 2}} \right. \kern-\nulldelimiterspace} 2}}e^{{\tau  \mathord{\left/
{\vphantom {\tau  2}} \right. \kern-\nulldelimiterspace} 2}}(-\,e^{-2\tau
}\,d\tau ^2+d\theta ^2)\nonumber \\
 &  &+e^{-\tau }\,[e^Pd\sigma ^2+2e^PQ\,d\sigma \,d\delta +(e^PQ^2+e^{-
P})\,d\delta ^2]
\end{eqnarray}
where $\lambda$, $P$, $Q$ are functions of $\theta$, $\tau$. We impose
$T^3$ spatial topology by requiring $0 \le \theta, \sigma, \delta \le 2 \pi$
and the metric functions to be periodic in $\theta$.
If we assume $P$ and $Q$ to be small, we find them to be respectively the
amplitudes of the $+$ and $\times$ polarizations of the gravitational
waves  with $\lambda$ describing the background in which they
propagate.  The time variable $\tau$ measures the area in the symmetry
plane with $\tau = \infty$ a curvature singularity. Einstein's equations split
into two groups. The first is nonlinearly coupled wave equations for $P$
and $Q$ (where $,_a = \partial / {\partial a}$):
\begin{eqnarray}
\label{gowdywave}
P,_{\tau \tau }-\;e^{-\kern 1pt2 \tau }P,_{\theta \theta }-e^{2P}\left(
{Q,_\tau ^2-\;e^{-\kern 1pt2\tau }Q,_\theta ^2} \right)&=&0,\\
  Q,_{\tau \tau }-\;e^{-\kern 1pt2\tau }Q,_{\theta \theta }+\,2\,\left(
{P,_\tau Q,_\tau ^{}-\;e^{-\kern 1pt2\tau }P,_\theta Q,_\theta ^{}}
\right)&=&0.
\end{eqnarray}
The second contains the Hamiltonian and $\theta$-momentum constraints
which can be expressed as first order equations for $\lambda$ in terms of
$P$ and $Q$:
\begin{equation}
\label{gowdyh0}
\lambda ,_\tau -\;[P,_\tau ^2+\;e^{-2\tau }P,_\theta ^2+\;e^{2P}(Q,_\tau
^2+\;e^{-2\tau }\,Q,_\theta ^2)]=0,
\end{equation}
\begin{equation}
\label{gowdyhq}
\lambda ,_\theta -\;2(P,_\theta P,_\tau +\;e^{2P}Q,_\theta Q,_\tau )=0.
\end{equation}
This break into dynamical and constraint equations removes two of the
most problematical areas of numerical relativity from this model:  (1) The
normally difficult initial value problem becomes trivial since $P$, $Q$
and their first time derivatives may be specified arbitrarily (as long as the
total $\theta$ momentum in the waves vanishes).  (2) The constraints,
while guaranteed to be preserved in an analytic evolution by the Bianchi
identities, are not automatically preserved in a numerical evolution with
Einstein's equations in differenced form.  However, in the Gowdy model,
the constraints are trivial since $\lambda$ may be constructed from the
numerically determined $P$ and $Q$.
For the special case of the polarized Gowdy model ($Q=0$), $P$ satisfies
a linear wave equation whose exact solution is well-known \cite{bkb74}.
For this case, it has been proven that the singularity is AVTD \cite{IM}.
This has also been conjectured to be true for generic Gowdy models
\cite{bm1}. We shall show in Section 4 how the SA applied to this model
provides strong support for this conjecture.

Our final model
generalizes the plane symmetry to a U(1) symmetry
\cite{Moncrief86}.
The details of this model will be given in
Section 5.  $U(1)$ models allow local Mixmaster dynamics
and thus can be used to test the BKL conjecture.  In fact, it is possible that
any type of allowed cosmological singularity will already appear in the
$U(1)$ models.  The extension of the Gowdy SA methods to the $U(1)$
case is straightforward.

\section{Symplectic Methods}
Consider the time evolution of a set of variables $X$ from $t_1$ to $t_2$.
We can define an evolution operator ${\cal U}(t_2,t_1)$ such that if
f $t_2-t_1 = \Delta t$ is infinitesimal, then ${\cal U}$ must have the form
\begin{equation}
\label{evolopinf}
{\cal U}(\Delta t) X = \left( 1 + \Delta t \frac{d}{dt} \right) X.
\end{equation}
But ${dX}/{dt} = \{H,X\}$ where $\left\{ {H,X} \right\}$
is the Poisson bracket with $H$ the Hamiltonian.  Thus ${\cal U}(\Delta
t) = 1 + \Delta t \{H, \quad \}$ with the empty slot in the operator to act on
$X$.  In the standard way (by dividing $\Delta t$ into $n$ intervals and
applying ${\cal U}({\Delta t}/n)$ $n$ times) we obtain the exponentiated
form (for finite $\Delta t$)
\begin{equation}
\label{evolop}
{\cal U}(\Delta t) = e^{ \Delta t \{H, \quad \}}  \equiv  e^{\Delta t A}.
\end{equation}
Suppose $H = H_1 + H_2$.  Then ${\cal U} = e^{\Delta t (A_1+A_2)}$.
Consider
\begin{equation}
\label{evop2nd}
{\cal U}_{(2)}(\Delta t)=e^{A_1(\Delta t/
2)}e^{A_2\Delta t}e^{A_1(\Delta t/ 2)} .
\end{equation}
Straightforward multiplication shows that
the right hand side of (\ref{evop2nd}) is a second order accurate
approximation to the evolution operator $e^{\Delta t A}$.  One evolves
$X$ from $t$ to $t + \Delta t$ by first evolving with $e^{{{(\Delta t}/2)}
A_1}$ from $t$ to $t + \frac{1}{2}{\Delta t}$, taking that result and
evolving with
$e^{\Delta t A_2}$ from $t$ to $t + \Delta t$, and, finally, evolving that
result with $e^{{{(\Delta t}/2)} A_1}$ from $t + \frac{1}{2}{\Delta t}$
to $t + \Delta t$.

Suzuki has shown how to obtain a representation of ${\cal U}$ to arbitrary
order \cite{suzuki}.  For example,
\begin{equation}
\label{evop4th}
{\cal U}_{(4)}(\Delta t)={\cal U}_{(2)}(s\Delta t){\cal U}_{(2)}[(1-
2s)\Delta
t]{\cal U}_{(2)}(s\Delta t)
\end{equation}
where $s=(2-2^{1/ 3})^{-1}$.  In general,
\begin{equation}
\label{evopmth}
{\cal U}_{(2m)}(\Delta t)={\cal U}_{(2m-2)}(s_m\Delta t){\cal U}_{(2m-
2)}[(1-2s_m)\Delta
t]{\cal U}_{(2m-2)}(s_m\Delta t)
\end{equation}
where $s_m=(2-2^{1/ {(2m-1)}})^{-1}$.  As a concrete example, consider
the Hamiltonian
\begin{equation}
\label{h1dof}
H = H_1 + H_2 = {\textstyle {1 \over 2}} p^2 +V(q)
\end{equation}
where $V$ is an arbitrary potential.  Note that both $H_1$ and $H_2$
yield exact solutions.  For $H_1$,
\begin{eqnarray}
\label{h1soln1dof}
q(t+\Delta t) &=& q(t) + p(t) \Delta t, \nonumber \\
p(t + \Delta t) &=& p(t),
\end{eqnarray}
while for $H_2$,
\begin{eqnarray}
\label{h2soln1dof}
q(t + \Delta t) &=& q(t), \nonumber \\
p(t + \Delta t) &=& p(t) - \left. \frac{d V}{d q} \right |_t \,
\Delta t .
\end{eqnarray}
Since $q$ is constant for the Hamiltonian $H_2$, the solution for $p$ is
exact no matter how complicated the potential $V$.  Thus the exact
solutions (\ref{h1soln1dof}) and (\ref{h2soln1dof}) are used to evolve
from $t$ to $t + \Delta t$ according to the prescription ${\cal
U}_{(2)}(\Delta t)$. To go to higher order, nothing new is required. The
time intervals are selected according to the prescription (\ref{evopmth}),
but the same exact solution is used.

Extension to fields $q(x,t)$, $p(x,t)$ is straightforward.  With computation
in mind, define $q_i^j \equiv q(x_i,t^j)$, etc.~where $x_i = i \Delta x$ and
$t^j = j \Delta t$. Then the exact solutions are for $H_1$
\begin{equation}
\label{h1soln1d}
( q_i^{j+1},p_i^{j+1} ) = ( q_i^j + p_i^j \Delta t, p_i^j )
\end{equation}
and for $H_2$
\begin{equation}
\label{h2soln1d}
\left( q_i^{j+1},p_i^{j+1} \right) = \left( q_i^j , p_i^j - \left. \frac{\delta
V}
{\delta q} \right |_i^j \Delta t \right)
\end{equation}
where ${\delta V / \delta q}$ is the appropriate functional derivative.
Again, the exact solution exists for $H_2$ no matter how complicated the
potential.  Of course, one must represent the spatial derivative ${\delta V /
\delta q}$ as accurately as possible.

To represent spatial derivatives to the desired order, evaluate (e.g.~in 1D)
the Taylor series for an arbitrary function $f(x)$.
Say the 4th order accurate expression for $d^2f/dx^2$ is desired.
Demand that
\begin{eqnarray}
\label{2nderiv}
a_1\left[ {f(x+\varepsilon )+f(x-\varepsilon )} \right]&+&a_2\left[
{f(x+2\varepsilon )+f(x-2\varepsilon )} \right]\nonumber \\
  &=&{{d^2f} \over {dx^2}}\varepsilon ^2+O(\varepsilon ^6).
\end{eqnarray}
We find $a_1 = \frac{4}{3}$, $a_2 = - \frac{1}{12}$.  A similar procedure
can be used
for any term in the Taylor expansion. It can also be extended to two spatial
dimensions where the same coefficients are found.  Extension to higher
order requires $f(x + n \varepsilon)$ for some $n > 2$ (depending on
order).

The only remaining issue in the evaluation of $( {{{dV}
\mathord{\left/ {\vphantom {{dV} {dq}}} \right. \kern-
\nulldelimiterspace} {dq}}} )_i^j$ is whether to vary
$V$ analytically and then difference the variation or difference $V$
and then vary the
differenced form.  We have chosen the latter.  (For complicated $V$, the
two are equivalent only to some order.) As a simple example, again
consider $d^2 f/dx^2$ to arise from the variation of
$V={1 \over 2}\left( {{{df} \over {dx}}} \right)^2$.
Since the Hamiltonian for this case is $H_2=\int {V[q(x,t)]\kern 1pt\,dx}$,
the differenced form is (for 2nd order accuracy)
\begin{equation}
H_2={1 \over 2}\sum\limits_{i=1}^N {{{\left( {f_{i+1}-f_i} \right)^2}
\over {(\Delta x)^2}}}.
\end{equation}
(The negative of) the variation with respect to $f_i$ (which appears in two
terms of the sum) yields the standard differenced second derivative
$(f_{i+1}+f_{i-1}-2 f_i)/(\Delta x)^2$.

\section{Mixmaster Model}
Although I have done several computational projects with the Mixmaster
model,
(see references in \cite{bkb94,bkb91}), none of these have used
the SA. Here we shall use Mixmaster primarily as an application of SA
rather than to evaluate Mixmaster parameters or to compute Lyapunov
exponents \cite{bkb90,bkb91}.  The Hamiltonian (\ref{mixh0}) clearly is
in the correct form with  $H_1=-p_\Omega ^2+p_+^2+p_-^2$ and $H_2
= U(\beta_{\pm}, \Omega)$ having obvious exact solutions.

(The variables chosen are not the only ones that have been used. For
example, the ADM reduction can be performed by choosing $\Omega$ to
be the time variable and solving the Hamiltonian constraint (\ref{mixh0}),
${\cal H} = 0$, for $p_{\Omega}$.  The remaining degrees of freedom for
$\beta_{\pm}$ are dynamical. Since the constraint is automatically
preserved, one might argue that this choice is better. However, the ability
to monitor $H$ as an indicator of the accuracy and stability of the solution
outweighs the disadvantage of the extra degree of freedom.)

\begin{figure}[bth]
\begin{center}
\setlength{\unitlength}{1cm}
\makebox[11.7cm]{\psfig{file=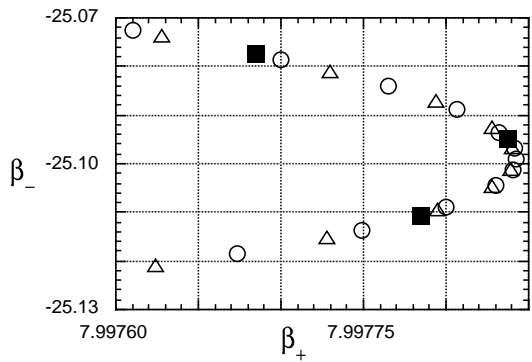,width=7cm}}
\caption[Comparison of SA and RKA]
{\protect \label{savsrk}
Close up of a Mixmaster bounce in the anisotropy
plane.  The same input file was evolved with a 4th order RKA
(open circles) and two SA's of 4th (open triangles) and 6th (filled squares)
order.  The same adaptive step size subroutine was used in
all cases.}
\end{center}
\end{figure}

There is actually some debate over the optimal choice of ODE solver (see
e.g.~\cite{nr}). The choice will involve trade-offs between accuracy and
computer time.  Accuracy is especially important for models with
sensitivity to initial conditions since a small numerical error could
qualitatively change the solution. (Sensitivity to initial conditions means
that an arbitrarily small change in initial data can generate qualitatively
different trajectories.  Numerical error could simulate change
in initial
conditions.) Here we shall consider only comparisons between a 4th order
Runge-Kutta algorithm (RKA) \cite{garcia} and 4th and 6th order SA's.
In all cases, the same initial data set (freely specify $\Omega$,
$\beta_{\pm}$, and $p_{\pm}$ and solve ${\cal H} = 0$ for $p_{\Omega}$) is
run for up to 2000 time steps. An adaptive step size algorithm has been
borrowed from \cite{garcia} for use in all these codes. A modification to
limit the maximum step size is required to avoid seriously over-running a
Mixmaster wall. When this eventually occurred anyway, the computation
was stopped. Fig.~\ref{savsrk}
displays the results for the three codes at a single
Mixmaster bounce.
Note that the 4th order SA and RKA have similar
performance while the 6th order SA equally well represents the solution
with many fewer time steps.  The 6th
order SA easily reached $\tau > 10^6$, more than two orders of magnitude
greater than RKA in essentially the same number of steps.
The much larger time step possible with the 6th order SA easily overcomes
tje extra computational time needed to take a 6th order step as three
4th order ones. It is only now
that SA's are being applied to the study of Mixmaster \cite{bkb95} and
related models \cite{sarwphi}.

\section{Gowdy Model on $T^3 \times R$}
The SA can be applied to the Gowdy cosmology
because the wave
equations (\ref{gowdywave}) can be obtained by variation of the
Hamiltonian
\begin{eqnarray}
\label{gowdywaveh}
H&=&{1 \over 2}\int\limits_0^{2\pi } {d\theta
\,\left[ {\pi _P^2+\kern 1pt\,e^{-2P}\pi _Q^2} \right]}\nonumber \\
  &+&{1 \over 2}\int\limits_0^{2\pi } {d\theta \,\left[ {e^{-
2\tau }\left( {P,_\theta ^2+\;e^{2P}Q,_\theta ^2} \right)}
\right]}=H_1+H_2.
\end{eqnarray}
Again equations obtained from the separate variations of $H_1$ and
$H_2$ are exactly solvable.  Variation of $H_1$ yields terms in
(\ref{gowdywave}) containing time derivatives.  These have the exact
(AVTD) solution
\begin{eqnarray}
\label{avtdeq}
P&=&-\beta \tau +\ln [\alpha \,(1+\zeta ^2e^{2\beta \tau })]\to \beta
\tau \quad {\rm as} \ \tau \to \infty ,\\
Q&=&-\;{{\zeta \,e^{2\beta \tau }} \over {\alpha \,(1+\zeta
^2e^{2\beta \tau })}}+\xi \quad\to Q_0\ \;{\rm as} \ \tau \to \infty
\ ,\\
\pi_P &=& {-{\beta (1-\zeta^2 e^{2 \beta \tau})} \over {(1-\zeta^2 e^{2 \beta
\tau})} } \quad \to \beta \ \; {\rm as} \ \tau \to \infty \ , \\
\pi_Q &=& -2 \alpha \beta \zeta
\end{eqnarray}
in terms of four constants $\alpha$, $\beta$, $\zeta$, and $\xi$.
To employ the AVTD solution in the SA, the values of
$P$, $Q$, $\pi_P$, and $\pi_Q$ at $t^j$ are used to find $\alpha$, $\beta$,
$\zeta$, and $\xi$.  These are substuted in (\ref{avtdeq}) to evolve to new
values at $t^{j+1}$ according to (\ref{evop2nd}). Note that
evolution with $H_1$ is purely local since there are no spatial derivatives.
This is advantageous for parallel processing.

Evolution with $H_2$ is still easy because $P$ and $Q$ are constant. The
necessary differencing has already been discussed in Section 2. For
completeness, we give the (2nd order) differenced form of $H_2$ as
\begin{equation}
\label{h2diffgowdy}
H_2={{e^{-2\tau }} \over {(\Delta \theta
)^2}}\sum\limits_{i=0}^N { {\left[ {\left(
{P_i-P_{i-1}} \right)^{2.}+e^{P_i+P_{i-1}}\left( {Q_i-Q_{i-
1}} \right)^2} \right]} .}
\end{equation}
The presence of only points and nearest neighbors in (\ref{h2diffgowdy})
also yields easy parallelization.  The exponential prefactor $e^{-2 \tau}$
in $H_2$ makes plausible the conjectured AVTD singularity.  However, $P
\to v \tau$ (for $v > 0$) as $\tau \to \infty$, (where from (\ref{avtdeq})
$v = \beta$). If $v > 1$, the term $e^{- 2 \tau} e^{2 P} Q,_{\theta}^2$ in
(\ref{gowdywaveh}) can grow rather than decay as $\tau \to \infty$. This
has led to the conjecture that the AVTD limit requires $v < 1$ everywhere
except, perhaps, at a set of measure zero (isolated values of $\theta$)
\cite{bm1}.

Implementation of the SA is completely straightforward. The published
results \cite{bkbvm} are based on a code which is 4th order accurate in
both time and space.  Actually, accurate results rely most strongly on
accurate representation of the spatial derivatives.  The time step must be
chosen to satisfy a Courant condition.
To test the code, we note that
it is possible to transform an exact solution $P_0 = Y_0
\left(e^{- \tau}\right) \cos \theta$ of the polarized case
(for $Y_n(x)$ an irregular Bessel function of order $n$) into the
pseudo-unpolarized solution $P = \ln \cosh P_0$, $Q = \tanh P_0$
which satisfies the full equations (\ref{gowdywave}).  Substitution shows
that the nonlinear terms (which must be absent in a polarized model)
miraculously cancel out.  However, the code is unaware of this {\em a
priori} and all terms are present.  The results are published elsewhere
\cite{bkbvm} and demonstrate the need for the 4th order code.

Most of the remaining results \cite{bkbvm,bggm}
use the initial data $P = 0$, $\pi_P = v_0
\cos \theta$, $Q = \cos \theta$, and $\pi_Q = 0$.  This model is actually
generic for the following reasons:  The $\cos \theta$ dependence is the
smoothest nontrivial possibility.  With $\cos n \theta$, the solution is
repeated $n$ times on the grid yielding the same result with
poorer resolution.  The amplitude of $Q$
is irrelevant since the Hamiltonian (\ref{gowdywaveh}) is invariant under
$Q \to \rho Q$, $P \to P - \ln \rho$.  This also means that any unpolarized
model is qualitatively different from a polarized ($Q = 0$) one no matter
how small $Q$.

\begin{figure}[bth]
\begin{center}
\setlength{\unitlength}{1cm}
\makebox[11.7cm]{\psfig{file=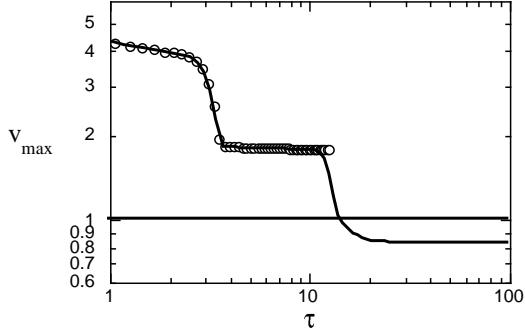,width=7cm}}
\caption[$v_{max}$ vs $\tau$]
{\protect \label{vmax}
Plot of $v_{max}$ vs $\tau$. The maximum value of $v$ is found for two
simulations with 3200 (solid line) and 20000 spatial grid points (circles)
respectively. The horizontal line indicates $v = 1$.}
\end{center}
\end{figure}

The accuracy and stability of the code easilty allow verification of the
conjectured AVTD behavior \cite{bkbvm}. A plot of the maximum value of
$v$ vs $\tau$ (Fig.~\ref{vmax}) shows strong support for the conjecture
that $v < 1$ in the AVTD regime.
However, Fig.~\ref{vmax} also shows that a simulation at higher spatial
resolution begins
to diverge. Normally, failure of convergence signals numerical problems. Here
something different is occurring. The evolution of spatial structure in $P$
depends on competition between the two nonlinear terms in the $P$ wave
equation. Approximating the wave equation by $P,_{\tau \tau} +
{\rm either \  of\  the\  nonlinear\  terms} = 0$, a first integral
can be obtained. The two potentials are $V_1 = \pi_Q^2 e^{-2P}$ and $V_2 =
e^{-2 \tau} e^{2P} Q,_{\theta}^2$.  Non-generic behavior can arise at
isolated points where either $Q,_{\theta}$ or $\pi_Q$ vanishes. Say such a
point is $\theta_0$. The finer the grid, the closer will be some grid point
to $\theta_0$. Thus non-generic behavior will become more visible on a finer
grid. Detailed examination shows that the differences seen in Fig.~\ref{vmax}
are due
to the slower decay of $v$ to a value below unity at isolated grid points in
the higher resolution simulation.

\begin{figure}[bth]
\begin{center}
\setlength{\unitlength}{1cm}
\makebox[11.7cm]{\psfig{file=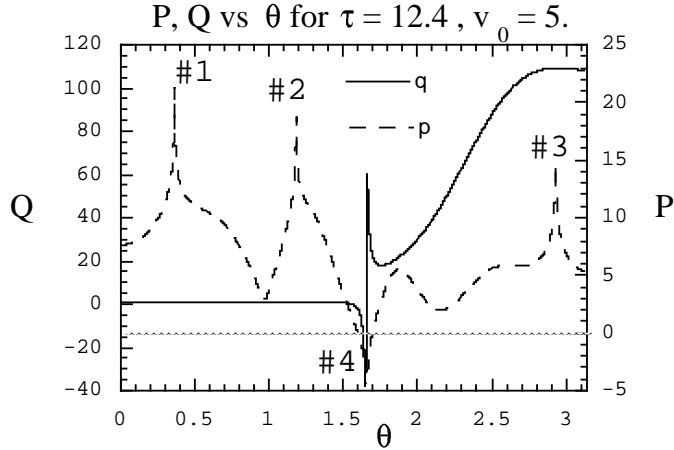,width=9cm}}
\caption[Gowdy $P$ and $Q$ Details]
{\protect \label{gdetail}
$P$ (dashed line) and $Q$ (solid line) vs $\theta$ at $\tau = 12.4$ for
the standard initial data set with $v_0 = 5$ for $0 \le \theta \le \pi$ for
a simulation containing 20000 spatial grid points in the interval $[0,2 \pi]$.
The numbers on the graph refer to the most interesting features. Peaks 1, 2,
and 3 in P are essentially the same in that they occur
where $Q,_{\theta} \approx 0$.
of $Q$. Peak 4 shows an apparent discontinuity in $Q$ where
$\pi_Q \approx 0$.}
\end{center}
\end{figure}

Details of the high resolution simulation (for $0 \le \theta \le 2 \pi$) are
shown in Fig.~\ref{gdetail}.
The narrow peaks in $P$ occur where
$Q,_{\theta} \approx 0$.
Generically, if $P \approx v \tau$ and $v > 1$, the potential $V_2$
dominates.
The relevant
first integral of (\ref{gowdywave}) is
\begin{equation}
\label{Zintegral}
\left( \frac {dZ}{d \tau} \right)^2 + e^{2Z} Q,_{\theta}^2 = const
\end{equation}
where $Z = P - \tau$. A bounce off $V_2$ yields ${dZ}/{d \tau}  \to -{dZ}/{d
\tau}$ or
$v - 1 \to 1-v$. If the new $v <0$, then $V_1$ dominates yielding the first
integral
\begin{equation}
\label{Pintegral}
P,_{\tau}^2 + e^{-2P} \pi_Q^2 = const
\end{equation}
causing $P,_{\tau} \to - P,_{\tau}$ or $v \to -v$.  Eventually, bouncing
between
potentials gives $v < 1$. However, if $Q,_{\theta} \approx 0$, but is not
precisely zero,
it takes a long time for the bounce off $V_2$ to occur.  Precisely at
$\theta_0$
(where $Q,_{\theta} = 0$), $v > 1$ persists.
The apparent discontinuity in $Q$ is not a numerical artifact. It occurs where
$P < 0$
and $\pi_Q \approx 0$. Since $Q,_{\tau} = e^{-2P} \pi_Q$ and $\pi_Q \approx c
(\theta
- \theta_0)$ (if $\pi_Q = 0$ at $\theta_0$), $Q,_{\tau}$ grows exponentially in
opposite directions about $\theta_0$. The potential $V_1$ drives $P$ to
positive
values unless $\pi_Q = 0$. Thus this feature will narrow as the simulation
proceeds.

\begin{figure}[bth]
\begin{center}
\setlength{\unitlength}{1cm}
\makebox[11.7cm]{\psfig{file=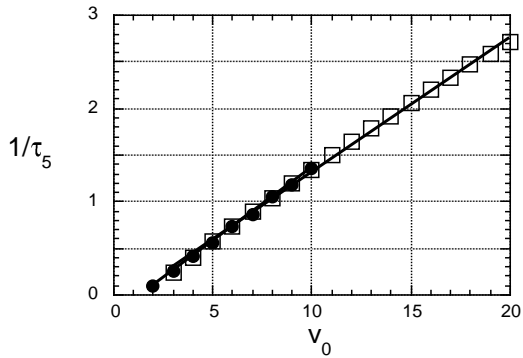,width=7cm}}
\caption[Scaling in the Gowdy Model]
{\protect \label{gscale}
Scaling in the Gowdy model. Plot of $1 / \tau_5$, the inverse of the time at
which
the 5th peak appears in $P$ vs $v_0$. Two cases are shown for initial data $P =
0$,
$Q = \cos \theta$: (1) $\pi_P = (v_0 / \protect \sqrt{2}) \cos \theta$, $\pi_Q
= (v_0 /
\protect \sqrt{2}) \cos \theta$ is indicated by filled circles; (2) $\pi_P =
v_0 \cos
\theta$, $\pi_Q = 0$ is indicated by open squares.}
\end{center}
\end{figure}

Finally, we report a strange, and yet not understood,
scaling of spatial structure in $P$ with the parameter $v_0$ in the initial
data. In
our standard initial data set, greater values of $v_0$ lead to the appearance
of
additional spatial structure in a shorter time. The rate of structure formation
decreases and then stops as AVTD is approached. One may count the number
of peaks in $P$ (a peak is crossed if $(P_{i+1} - P_i)(P_i - P_{i-1}) < 0$ and
$P_{i+1} < P_i$) during the simulation. The scaling is best for the time
$\tau_5$
at which the 5th peak appears although it is also seen for $\tau_3$ and
$\tau_7$
(the even nature of the solution causes two peaks to appear at once except at
$\theta = \pi$). A plot of
$1/{\tau_5}$ vs $v_0$ yields a straight line as shown in Fig.~\ref{gscale}
which
may be described as $\tau_5 = a (v_0 - v_0^{\infty})^{-1}$
where, if $v_0 = v_0^{\infty}$, the 5th peak does not appear until $\tau =
\infty$.
Even more surprisingly, almost the same line is obtained for initial data
$\pi_P = ({v_0}/{\sqrt{2}}) \cos \theta$, $\pi_Q = ({v_0}/{\sqrt{2}}) \cos
\theta$
rather than our standard data. Explanation of this scaling is still in
progress.

\section{$U(1)$ models}
Moncrief has shown \cite{Moncrief86} that cosmological models on $T^3 \times R$
with a spatial $U(1)$ symmetry
can be described by five degrees of freedom
$\{ x,z, \Lambda, \varphi, \omega \}$ and their respective conjugate momenta
$\{ p_x, p_z, p_{\Lambda}, p, r \}$.  All variables are functions of spatial
variables $u$, $v$ and time $\tau$ (related to distance in the symmetry
direction).
If we define a conformal metric $g_{ab}$ in the $u$-$v$ plane as $g_{ab} =
e^{\Lambda} e_{ab}(x,z)$ where
\begin{equation}
\label{eab}
e_{ab} = \frac{1 }{ 2}  \left( \begin{array}{cc}
          e^{2z}+e^{-2z}(1+x)^2 & e^{2z}+e^{-2z}(x^2 - 1) \\
                                                           \\
           e^{2z}+e^{-2z}(x^2 - 1) &  e^{2z}+e^{-2z}(1-x)^2
         \end{array} \right)
\end{equation}
has unit determinant and choose the lapse $N = e^{\Lambda}$, Einstein's
equations can be obtained by variation of
\begin{eqnarray}
\label{u1h}
H&=&-\oint {\oint {du\,dv\left\{ {{\textstyle{1
\over 8}}p_z^2+{\textstyle{1 \over
2}}e^{4z}p_x^2+{\textstyle{1 \over 8}}p_{}^2+{\textstyle{1
\over 2}}e^{4\varphi }r_{}^2-{\textstyle{1 \over 2}}p_\Lambda
^2-2p_\Lambda } \right.}} \nonumber \\
& &+e^{-2\tau }\left[ {\left( {e^\Lambda e^{ab}} \right),_{ab}-
\;\left( {e^\Lambda e^{ab}} \right),_a\Lambda ,_b+\;e^\Lambda
\left( {e^{-2z}} \right),_{[a}x,_{b]}} \right. \nonumber \\
 & &\left. {\left. {+2e^\Lambda e^{ab}\varphi ,_a\varphi
,_b+{\textstyle{1 \over 2}}e^\Lambda e^{-4\varphi
}e^{ab}\omega ,_a\omega ,_b} \right]} \right\} \nonumber \\
  &=& H_1 + H_2 .
\end{eqnarray}
Note particularly that
\begin{equation}
\label{u1parts}
H_1 = H_1^G(-2 z,x)+ H_1^G(-2 \varphi, \omega)+H_1^F(\Lambda)
\end{equation}
where $H_1^G(P,Q)$ is the kinetic part of the Gowdy Hamiltonian
(\ref{gowdywaveh}).
Of course, $H_1^F$ is just a free particle Hamiltonian for the degree of
freedom associated with $\Lambda$. This means that not only are the
equations from $H_1$ exactly solvable but also that the Gowdy coding
can be used with essentially no change. The potential term $H_2$ is very
complicated. However, it still contains no momenta so its equations are
trivially exactly solvable. Thus, at least in principle, the extension
of the Gowdy code to the two spatial dimensions of the $U(1)$ code is
completely straightforward.

There are three complications, however, which cause the $U(1)$ problem
to be more difficult. The first involves the initial value problem
(IVP)---the constraints must be satisfied on the initial spacelike slice.
The constraints are
\begin{equation}
\label{u1h0}
{\cal H}_0 = {\cal H} - 2 p_{\Lambda} = 0
\end{equation}
(where ${\cal H}$ is the density in (\ref{u1h})) and
\begin{equation}
\label{u1ha}
H_a=-2 \tilde \pi^b_{a;b}+p_\Lambda \Lambda ,_a-
p_\Lambda ,_a+p\varphi ,_a+r\omega ,_a = 0
\end{equation}
where $\tilde \pi^b_{a}$ is in the 2-space with metric $e_{ab}$
and is linear in $p_x$ and $p_z$ with each term containing one or the other.
Moncrief has proposed a particular solution to the IVP. First, identically
satisfy
${\cal H} = 0$ by choosing
\begin{equation}
\label{ivp}
p_x=p_z= \varphi ,_a=\omega ,_a=0 \quad ; \quad p_{\Lambda}
= c \, e^{\Lambda}
\end{equation}
for $c$ a constant. Then, solve ${\cal H}^0 = 0$ for either $r$ or $p$.
Solution is
possible for $c \ge c_{min}$ such that $r^2$ or $p^2 \ge 0$. This allows $x$,
$z$, $\Lambda$, and $p$ or $r$ to be freely specified.
(Without loss of generality, it is possible to set $x = z= 0$ initially
to yield $e_{ab}$ flat. Such a condition may always be imposed at one time
by rescaling $u$ and $v$.)

The second difficulty also involves the constraints. While the Bianchi
identities guarantee the preservation of the constraints by the Einstein
evolution equations, there is no such guarantee for differenced evolution
equations. At this stage of the project, we monitor the maximum value of
the constraints vs $\tau$ over the spatial grid but do nothing else to try
to stay on the constraint hypersurface.

The third difficulty, and the one that is proving to be the greatest obstacle,
is instability associated with spatial differencing in two dimensions. The
differencing algorithm of section 2 for sixth order accurate expressions
(based on second order ones) is used. The difficulties we encounter are
apparently common to nonlinear equations in two or more dimensions. In an
attempt to control the instability, we have introduced a form of spatial
averaging. At the end of every time step, each value of every variable, say
$\xi_{ij}$, is replaced by
\begin{eqnarray}
\label{avg}
\bar \xi _{i,j}&= &{\textstyle{1 \over 2}}\xi
_{i,j}+{\textstyle{1 \over 4}}\left( {\xi _{i+1,j}+\xi _{i-
1,j}+\xi _{i,j+1}+\xi _{i,j-1}} \right) \nonumber \\
& & -{\textstyle{1 \over 8}}\left( {\xi _{i+1,j+1}+\xi _{i-1,j-
1}+\xi _{i+1,j-1}+\xi _{i-1,j+1}} \right).
\end{eqnarray}
This expression gives $\bar \xi_{ij} = \xi_{ij} + {\cal O}(\Delta^4)$.
In both test cases and generic models, the averaging procedure has
allowed the code to run longer.  However, the fact that $\bar \xi_{ij}
\ne \xi_{ij}$ can lead to deviations of the numerical solution from the
correct one. Fortunately, by comparing runs with and without averaging,
these artifacts are easy to identify. We shall see some examples of how
averaging can allow the code to run long enough for a conclusion about the
asymptotic singularity behavior to be drawn.

Moncrief has provided a test case for the $U(1)$ code. It again starts with
a polarized Gowdy solution and transforms it as either a 1D ($\theta \to u$ or
$v$)
or 2D ($\theta \to f(u,v)$) test problem to satisfy the $U(1)$
equations (including the constraints).
As a 1D example, the agreement is excellent and the code can be run to
large $\tau$. Difficulties arise in the 2D test problem in regions where the
spatial
derivatives are large.
In application of the $U(1)$ code to generic models, AVTD models
can display nonlinear wave interactions before settling down to $U \to
0$, $z,\varphi, \Lambda \to const \  \tau$, and $x,\omega \to const$.
Increasing spatial resolution will yield narrower (and steeper)
structures and thus may not help to cure instabilities due to steep
gradients.

\begin{figure}[bth]
\begin{center}
\setlength{\unitlength}{1cm}
\makebox[11.7cm]{\psfig{file=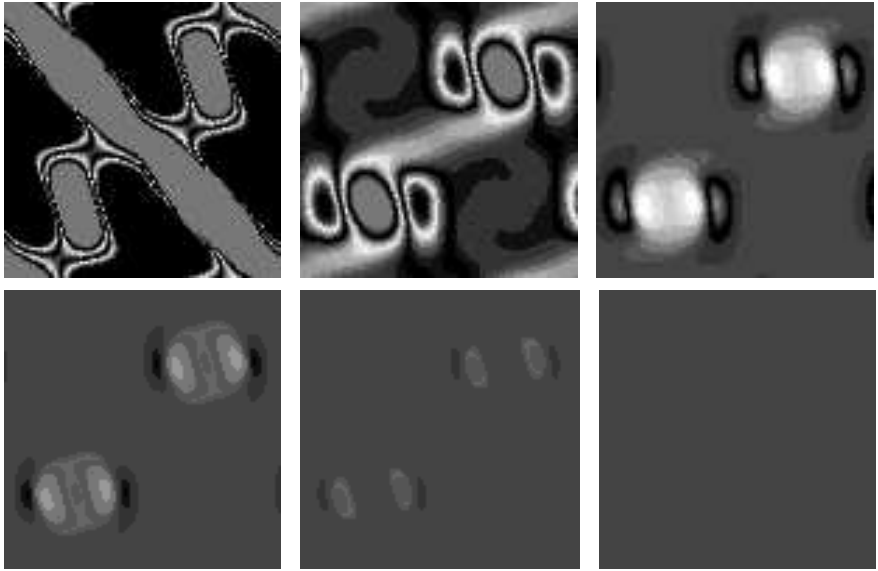,width=11.7cm}}
\caption[Frames of $U(u,v,\tau)$ for the Polarized Model]
{\protect \label{polU}
Frames of $U(u,v,\tau)$ for the
polarized model $x = z = \Lambda = \sin u \sin v, \  p_\Lambda = 12
e^{\Lambda}, \  \omega = r = 0$.  Time increases to the right and downward.
The final frame corresponds to $U \approx 0$ everywhere.}
\end{center}
\end{figure}

\begin{figure}[bth]
\begin{center}
\setlength{\unitlength}{1cm}
\makebox[11.7cm]{\psfig{file=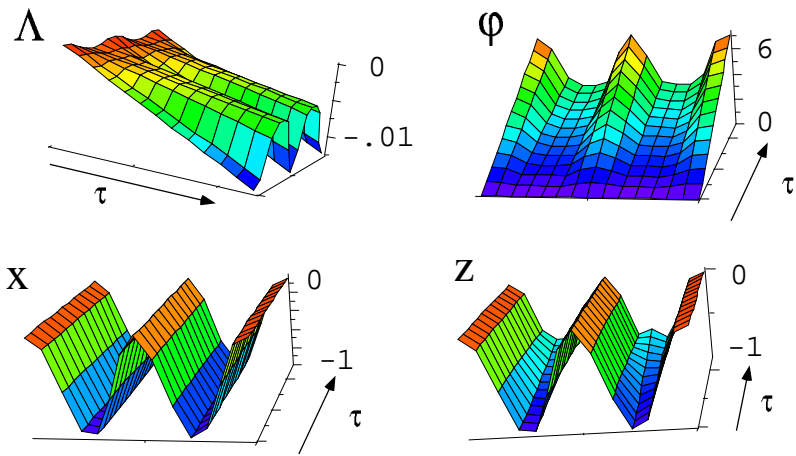,width=11.7cm}}
\caption[Surface Plot of All $U(1)$ Variables]
{\protect \label{polsurf}
Surface plot of all $U(1)$ variables for
the line $u = -v$ vs $\tau$ for the polarized model. Note agreement with
predicted AVTD behavior that $x$ becomes constant in $\tau$ while $\Lambda$,
$\varphi$, and $z$ grow linearly.}
\end{center}
\end{figure}

\begin{figure}[bth]
\begin{center}
\setlength{\unitlength}{1cm}
\makebox[11.7cm]{\psfig{file=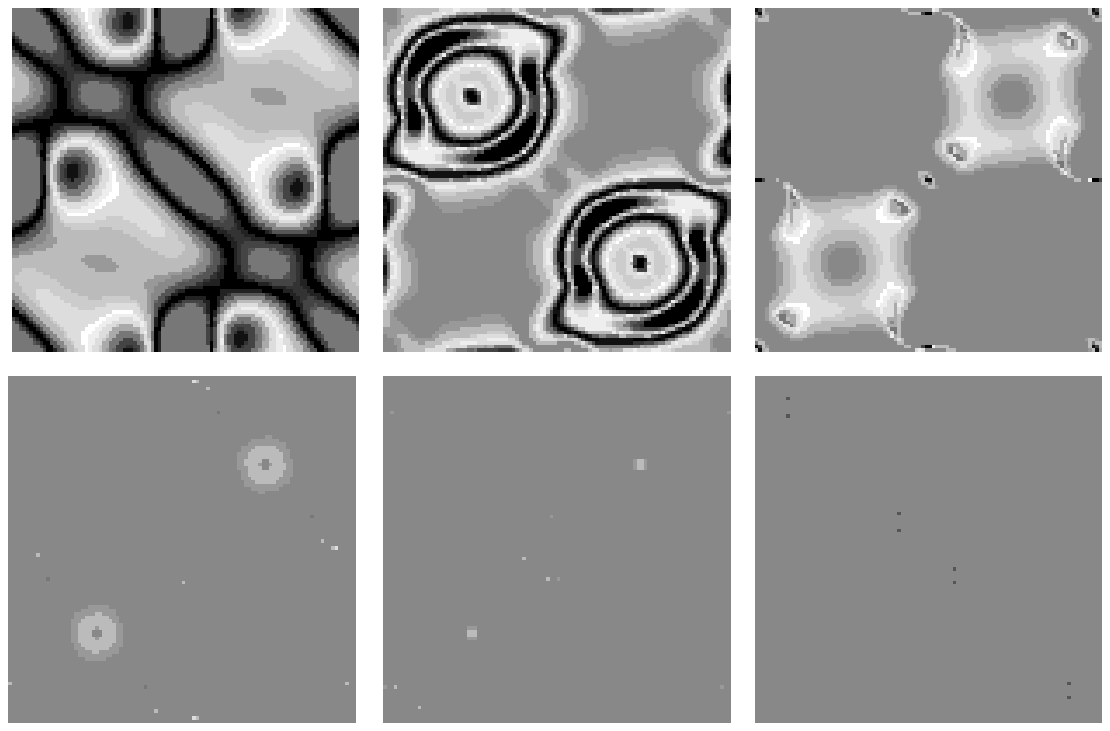,width=11.7cm}}
\caption[Frames of $U(u,v,\tau)$ for a Generic Model]
{ \protect \label{genU}
Frames of $U(u,v,\tau)$ for the generic model
$x = z = \cos u \cos v , \ \Lambda = \sin u \sin v, \ p_\Lambda =
14 e^{ \Lambda}, \  p = 10 \cos u \cos v$ with averaging.}
\end{center}
\end{figure}

\begin{figure}[bth]
\begin{center}
\setlength{\unitlength}{1cm}
\makebox[11.7cm]{\psfig{file=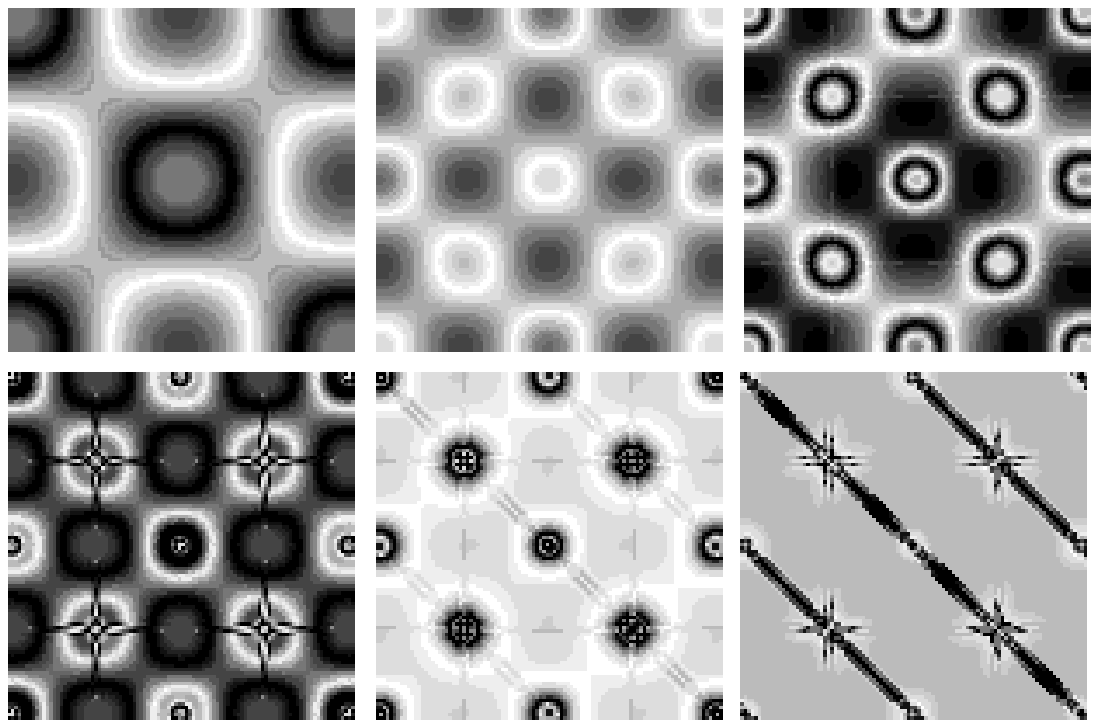,width=11.7cm}}
\caption[Frames of $U(u,v,\tau)$ for a Second Generic Model]
{\protect \label{genU2}
Frames of $U(u,v,\tau)$ for generic model $x = z = 0,
\  \Lambda = .1 \cos u \cos v, \  p_\Lambda= 2.1 e^
{\Lambda}, \  r = \cos u \cos v$. The diagonal features in the final frames
are numerical artifacts.}
\end{center}
\end{figure}

Nonetheless, conclusions can be drawn for generic models in our restricted
class of initial data. We shall consider the models as representative of
subclasses of the data. Models with $r = \omega = 0$ are called polarized. This
condition
is compatible with the above solution and is preserved (identically) by the
(analytic and numerical) evolution equations. Grubi\u{s}i\'{c} and Moncrief
have
conjectured that these polarized models are AVTD \cite{bm2}.
Therefore, the first model is chosen to be polarized. It exhibits the
conjectured AVTD behavior as shown in Figs.~\ref{polU} and \ref{polsurf}.
The second is generic with
$p$ given and the Hamiltonian constraint solved for $r$ in the IVP.
Averaging as in (\ref{avg}) allows the model to be followed to the point
where only artifacts have $U \ne 0$. This is shown in Fig.~\ref{genU}.
The last
model has $r$ given with $p$ obtained by solving the Hamiltonian constraint
in the IVP. Models of this type are less stable, probably due to the
growth of a steep feature in $\omega$ which does not appear in the other
cases. For this reason, the parameters must be kept small. Fig.~\ref{genU2}
shows
that $U \to 0$ except near artifacts where no statement can be made.

\section{Conclusions}
The application of the SA to Einstein's equations for collapsing universes
allows the nature of the singularity to be studied. While no real
attack on  Mixmaster with SA has been made, the method offers the potential
for efficient, accurate evolution of this model.
Application of SA to the Gowdy model has yielded strong support for its
conjectured AVTD singularity and has allowed the discovery and study
of interesting small scale spatial structure and scaling.

Further progress in understanding the generic singularity of $U(1)$
cosmologies
requires imporvements in handling steep spatial gradients.
Nevertheless there is strong support that (at least within our restricted
class of initial data) polarized models are AVTD. There is also support
for AVTD behavior in all generic models studied so far. Mixmaster-like
bounces have not been seen. (The activity due to nonlinear wave
interaction seen early in the simulations is similar to that in Gowdy
models.) Several factors could account for this: (1) the BKL conjecture
is false; (2) the simulations have not run long enough; (3) Mixmaster
behavior is present but hidden in our variables; or (4) our class of
initial data is insufficiently generic. All these possibilities will be
explored in studies in progress.

\section*{Acknowledgements}
I would like to thank the Astronomy Department of the University of Michigan,
the Institute for Geophysics and Planetary Physics at Lawrence Livermore
National Laboratory, and the Albert Einstein Institute at Potsdam for
hospitality. This work was supported in part by National Science Foundation
Grants PHY93-0559 and PHY9507313. Computations were performed at the National
Center for Supercomputing Applications (University of Illinois) and at the
Pittsburgh Supercomputer Center.

\begin{appendix}
\section*{Appendix}\setcounter{section}{1}
\addcontentsline{toc}{section}{Appendices}
This Appendix contains a FORTRAN subroutine which uses the 6th order SA to

integrate Einstein's equations for the Mixmaster universe:
\begin{equation}
{{d\Omega } \over {dt}}=-p_\Omega \quad,\quad{{d\beta _\pm } \over {dt}}=p_\pm
,
\end{equation}
\begin{equation}
{{dp_\Omega } \over {dt}}=-\,4\,e^{4\Omega }V(\beta _\pm )\quad,\quad{{dp_\pm }
\over {dt}}=-\,e^{4\Omega }{{\partial V(\beta _\pm )} \over {\partial \beta
_\pm
}}
\end{equation}
where
\begin{eqnarray}
V(\beta _\pm )&=&e^{-8\beta _+}+e^{4(\beta _++\sqrt 3\beta _-)}+e^{4(\beta
_+-\sqrt
3\beta _-)}\nonumber \\
  & &-2e^{4\beta _+}-2e^{-2(\beta _++\sqrt 3\beta _-)}-2e^{-2(\beta _++\sqrt
3\beta _-
)} .
\end{eqnarray}

\tt
{\parindent=1.5cm              subroutine sa6(x,t,tau,param,nstate,xout)}

{\parindent=0cm
$\ast$TAKES ONE 6TH ORDER SA STEP FROM T TO T + TAU

$\ast$

$\ast$VARIABLES: X(1) = BP, X(2) = BM, X(3) = W

$\ast$X(4) = PP, X(5) = PM, X(6) = PW

$\ast$

$\ast$XOUT CONTAINS THE OUTPUT VALUES OF THE VARIABLES

$\ast$PARAM = SQRT(3), NSTATE = 6

$\ast$

$\ast$XTEMP AND XS ARE DUMMY ARRAYS

$\ast$V IS THE POTENTIAL, DVDBP,DVDBM ARE THE GRADIENTS

$\ast$TERM\_I IS AN EXPONENTIAL

$\ast$

$\ast$S, S1 ARE THE 4TH,6TH ORDER SUZUKI PARAMETERS

$\ast$DELT,DELT1 STORE T INTERVALS}

{\parindent=1.5cm               parameter(mstate=20)

                real$\ast$8 x(6),param(1),xout(6),xtemp(6),xs(6)

                real$\ast$8 term1,term2,term3,term4,term5,term6,}

{\parindent=1cm     \&v,dvdbp,dvdbm,a,s,delt(3),s1,delt1(3)}

{\parindent=0cm
$\ast$

$\ast$DOUBLE PRECISION IS NECESSARY

$\ast$S WAS COMPUTED WITH MATHEMATICA  }

{\parindent=1.5cm
                 s=1.35120719195965

                 s1=1.d0/(2.d0-2.d0$\ast$$\ast$(.2d0))}

{\parindent=0cm
$\ast$

$\ast$STORE THE VARIABLES IN A DUMMY ARRAY}

{\parindent=1.5cm                do i=1,6

                xs(i) = x(i)

                enddo}

{\parindent=0cm
$\ast$

$\ast$ONE 6TH ORDER STEP IS THE PRODUCT OF 3

$\ast$4TH ORDER STEPS

$\ast$

$\ast$COMPUTE THE INTERVALS FOR THE 4TH ORDER STEPS}

{\parindent=1.5cm
                delt1(1) = s1$\ast$tau

                delt1(2) = (1.d0-2.d0$\ast$s1)$\ast$tau

                delt1(3) = s1$\ast$tau}

{\parindent=0cm
$\ast$THE DELT1 SUM TO TAU

$\ast$

$\ast$EACH 4TH ORDER STEP IS THE PRODUCT OF 3

$\ast$2ND ORDER STEPS

$\ast$}

{\parindent=1.5cm                do idel1=1,3}

{\parindent=0cm
$\ast$

$\ast$EACH 4TH ORDER INTERVAL IS DIVIDED INTO THREE

$\ast$INTERVALS FOR 2ND ORDER

$\ast$}

{\parindent=1.5cm
                delt(1) = s$\ast$delt1(idel1)

                delt(2) =(1.d0-2.d0$\ast$s)$\ast$delt1(idel1)

                delt(3) = s$\ast$delt1(idel1)}

{\parindent=0cm
$\ast$}

{\parindent=1.5cm                do idel = 1,3}

{\parindent=0cm
$\ast$

$\ast$PERFORM A SECOND ORDER SA STEP FOR EACH INTERVAL

$\ast$   }

{\parindent=1.5cm
                half\_tau=0.5$\ast$delt(idel)

                a = param(1)}

{\parindent=0cm
$\ast$

$\ast$EVOVLE WITH H\_1 FOR HALF A TIME STEP

$\ast$}

{\parindent=1.5cm
                xtemp(1) = xs(1)+xs(4)$\ast$half\_tau

                xtemp(2) = xs(2)+xs(5)$\ast$half\_tau

                xtemp(3) = xs(3)-xs(6)$\ast$half\_tau}

{\parindent=0cm
$\ast$

$\ast$EVALUATE THE EXPONENTIALS IN THE MOST ACCURATE WAY    }

{\parindent=1.5cm
                term1=exp(4$\ast$xtemp(3)-8$\ast$xtemp(1))

                term2=exp(4$\ast$xtemp(3)+4$\ast$xtemp(1))

term3=exp(4$\ast$xtemp(3)+4$\ast$xtemp(1)+4$\ast$a$\ast$xtemp(2))

term4=exp(4$\ast$xtemp(3)-2$\ast$xtemp(1)-2$\ast$a$\ast$xtemp(2))

term5=exp(4$\ast$xtemp(3)+4$\ast$xtemp(1)-4$\ast$a$\ast$xtemp(2))

term6=exp(4$\ast$xtemp(3)-2$\ast$xtemp(1)+2$\ast$a$\ast$xtemp(2))
}

{\parindent=0cm
$\ast$

$\ast$COMPUTE V AND ITS GRADIENTS     }

{\parindent=1.5cm
                v=term1+term3+term5-2.$\ast$(term2+term4+term6)

                dvdbp=4.$\ast$(-2.$\ast$(term1+term2)+term3+term4}

{\parindent=1cm     \&+term5+term6)}

{\parindent=1.5cm
                dvdbm=4.$\ast$a$\ast$(term3+term4-term5-term6)}

{\parindent=0cm
$\ast$

$\ast$EVOLVE WITH H\_2 FOR A FULL TIME STEP   }

{\parindent=1.5cm
                xs(4) = xs(4) - .5$\ast$dvdbp$\ast$delt(idel)

                xs(5) = xs(5) - .5$\ast$dvdbm$\ast$delt(idel)

                xs(6) = xs(6) - 2.$\ast$v$\ast$delt(idel) }

{\parindent=0cm
$\ast$

$\ast$EVOLVE WITH H\_1 FOR HALF A TIME STEP

$\ast$}

{\parindent=1.5cm
                xs(1) = xtemp(1) + xs(4)$\ast$half\_tau

                xs(2) = xtemp(2) + xs(5)$\ast$half\_tau

                 xs(3) = xtemp(3) - xs(6)$\ast$half\_tau}

{\parindent=0cm
$\ast$}

 {\parindent=1.5cm
               enddo

                enddo}

{\parindent=0cm
$\ast$

$\ast$RECORD THE DUMMY ARRAY   }

{\parindent=1.5cm
                do i=1,6

                xout(i) = xs(i)

                enddo}

{\parindent=0cm
$\ast$}

 {\parindent=1.5cm               return

                end}
\bigskip

{\parindent=0cm
OTHER CODE FEATURES:

Input w, bp, bm, pp, pm and solve H for pw.

A pos\-i\-tive square
root yields a col\-lap\-sing uni\-verse.

Use an adap\-tive step size but require tau $\le$ .01 t. }

\end{appendix}

\end{document}